# Neural development features: Spatio-temporal development of the *Caenorhabditis elegans* neuronal network

Running Title: **Neural development of *Caenorhabditis elegans***

Sreedevi Varier[1] and Marcus Kaiser[1,2,3] *

* M. Kaiser@ncl.ac.uk

[1] School of Computing Science, Newcastle University, Claremont Tower, Newcastle upon Tyne, NE1 7RU, United Kingdom
[2] Institute of Neuroscience, Henry Wellcome Building for Neuroecology, University of Newcastle, Framlington Place, Newcastle upon Tyne, NE2 4HH, United Kingdom
[3] Department of Brain and Cognitive Sciences, Seoul National University, Shilim, Gwanak, Seoul 151-746, Korea

**Abstract**

The nematode *Caenorhabditis elegans*, with information on neural connectivity, three-dimensional position and cell linage provides a unique system for understanding the development of neural networks. Although *C. elegans* has been widely studied in the past, we present the first statistical study from a developmental perspective, with findings that raise interesting suggestions on the establishment of long-distance connections and network hubs. Here, we analyze the neuro-development for temporal and spatial features, using birth times of neurons and their three-dimensional positions. Comparisons of growth in *C. elegans* with random spatial network growth highlight two findings relevant to neural network development.

First, most neurons which are linked by long-distance connections are born around the same time and early on, suggesting the possibility of early contact or interaction between connected neurons during development. Second, early-born neurons are more highly connected (tendency to form hubs) than later born neurons. This indicates that the longer time frame available to them might underlie high connectivity. Both outcomes are not observed for random connection formation. The study finds that around one-third of electrically coupled long-range connections are late forming, raising the question of what mechanisms are involved in ensuring their accuracy, particularly in light of the extremely invariant connectivity observed in *C. elegans*.

In conclusion, the sequence of neural network development highlights the possibility of early contact or interaction in securing long-distance and high-degree connectivity.

**Author Summary**

Long-distance connections are crucial for information processing in neural systems and changes in long-distance connectivity have been shown for many brain diseases ranging from Alzheimer's to schizophrenia. How do long-distance connections develop? Traditionally, connections can be formed over long distances using guidance cues for steering axonal growth. Subsequently, other connections can follow those pioneer axons to a target location. Alternatively, two neurons can establish a connection early on which turns into a long-distance connection as the neural system grows. However, the relative contribution of both mechanisms previously remained unclear. Here, we study long-distance connection development in the neuronal network of the roundworm *C. elegans*. We find that most neurons that are connected by a long-distance connection could interact and establish contact early on. This suggests that early formation could be an influential factor for establishing long-distance connectivity, with a hypothetical role in neuronal wiring accuracy. Reducing the need for axonal guidance is also likely to reduce metabolic costs during development. We also find that highly-connected neurons (hubs) are born early on, potentially giving them more time to host and establish connections. Therefore, neuron birth times can be an important developmental factor for the spatial and topological properties of neural circuits.

**Introduction**

The complexity of the nervous system continues to protract efforts to understand its development. The relatively simple invertebrate neural systems have been the subject of intense study in the last few decades [1], helping to shed light on mechanisms involved in development like axon guidance and molecular cues. We looked at the development of the neuronal network of *C. elegans* using information from 279 of the 302 neurons (see methods and [2]). Information on embryonic and post-embryonic lineages of the neurons [3,4,5,6] formed the basis of our developmental data.

Our work marks the first attempt to computationally and statistically represent the neural development of *C. elegans* based on available biological data, enabling a spatio-temporal analysis of the developing neuronal network. Graph theory [7] is increasingly being applied to elucidate the function based on the structures of complex networks like the brain [8,9] and here we carry out a structural analysis of neuronal networks during different stages of *C. elegans* development. We observe neuronal growth (see Video S1), development times of different classes of neurons, and the time windows for establishing short- and long-distance connectivity.

It is now known that *C. elegans* displays a higher than expected wiring cost with the total wiring length of all connections being twice as high as for an optimized network with spatially rearranged neurons [2,10]. Instead, processing speed indicated by the number of intermediate neurons in a pathway and made possible by long-distance connections, seems to be a critical constraint [2]. Establishing accurate connectivity, particularly long-distance, is a critical challenge in development. While the role of guidance molecules is well established in correctly wiring neurons, our results suggest that temporal and possible spatial closeness permitting early neuron-neuron interaction could also have an important role to play in the process. It may be conjectured that this would serve to reduce metabolic costs during development and also increase the probability of accurately establishing long-distance connections. Long-distance connections are vital and known to be affected in several neurological disorders in humans [11].

**Results**

Our analysis was primarily four-fold. First, we traced the time course of neuron establishment. Second, we looked at temporal patterns emerging from birth times of neurons and their known connectivity details. Third, we looked at spatial network features in particular focusing on the onset of short, medium, and long distance connection-pairs. The analysis was further refined by segregating connections based on the nature of transmission (i.e. gap junctions or chemical synapses) as well as their membership in functional circuits. Finally, we analyzed possible networks at the various stages identified.

**Time course of neuron formation**

The growth in the neuronal network was visualized with respect to the spatial positions of neurons in the nematode body. In the development phase, neurons are born in two bursts – a relatively brief embryonic burst lasting around four hours and a longer post-embryonic phase stretching across seventeen hours. These intervals have been sub-divided into smaller time intervals to enable a more detailed visualization of the sequential appearance of neurons during these bursts. In Figure 1 neurons appearing at a particular stage are indicated in black, while previously existing neurons are coloured in gray. In the first stage, at the end of 350 minutes of embryonic growth, neurons that eventually reside in the head, along the ventral cord (classes DAn, DBn and DDn) and tail of the nematode have appeared, with neurons in the head constituting the majority. By the end of the embryonic burst nearly all of the neurons in the head have appeared. Ventral cord neurons belonging to the other five classes namely, ASn, VAn, VBn, VCn and VDn as well as the rest of the tail neurons are born in the latter phase, towards the end of the first larval stage. A relevant question is whether the spatial clustering of neurons in the head, body and tail of the worm also relates to their topology. We find that neurons in the head are mostly connected to neurons in the head (74% of all connections) while 46% of connections of neurons in the body are to other neurons also in the body. Within spatial cluster connectivity is least for neurons in the tail at 29%. Figure S1 shows the successive appearance of neurons with respect to their final 3-dimensional positions in the adult body on a colour scale.

**Analysis of temporal features**

We plotted a histogram of the differences in birth-times of connected neurons to assess the time interval available for formation of connections (Figure 2A). For comparison, we looked at the outcomes from twenty random networks (see Methods for details). The values from the trials were separated into ten time bins and the mean and standard deviation values were calculated for each of the bins. It can be seen that approximately two-thirds of connected neurons appear less than 200 minutes apart and this proportion is much higher than that seen in random networks. A zoom-in on time differences of up to 500 minutes shows that most connected neurons within this interval are born less than 50 minutes apart (Figure 2B).

Neurons appearing in the embryonic burst go on to make most of their connections, approximately 80%, with other neurons born in the same phase, while this figure is around 46% for neurons born in the post-embryonic burst. These values in randomly shuffled networks are on an average 62.1% and 38.2%, for the embryonic and post-embryonic phases, respectively. This observation could well be attributed to high degree neurons being born in the embryonic phase (discussed below), however a statistical comparison of the ratio of connections made within and outside the temporal cluster for the embryonic and post-embryonic phases to random development, highlights significance with $p$ value less than 0.001 in a one sample t-test.

Early stages are also crucial for the formation of highly connected neurons (hubs). Figure 3 shows node degree (the number of connections of a neuron) with respect to birth times of neurons and their positions along the body of the worm, namely head, body or tail. Neurons that appear early on in development are more likely to have a higher degree than those that are born later. As would be expected, the highest degree is possessed by neurons in the head. There are 112 neurons with 20 or more connections. This number drops sharply to 46 and then to 9, for degrees more than or equal to 30 and 60, respectively. Of all neurons with degree 20 or more, two thirds appear before hatching (840 minutes), and nearly all neurons (~98%) with more than 30 connections are born before hatching. It needs to be noted however, that correlation of early birth and higher degree is less obvious for neurons with less than 30 connections. To gauge the significance of these results we repeated the test for twenty random networks (details in Methods). For this random connection formation, approximately 74% (±5.5) of neurons with degrees above 30 could appear before hatching, which using two-sided t-test corresponded to a *p* value of less than 0.001 indicating significant difference between actual and random networks. We also analyzed how the connected neighbors of a neuron appeared over time, particularly to examine if longer time windows served to receive connections from late appearing neurons. This was indeed the case: all neurons with a degree of more than 60 were connected to more than one late forming neighbour. However in random networks, this was on average true for 60% of the cases (based on 20 random observations) and had a statistically lower connectivity with late appearing neurons, with $p = 0.001$. Thus availability of time between neuron birth and nematode maturation appears to be important to hub neurons.

The time differences of bilaterally paired neurons were also compared for examining symmetry during development. All bilateral pairs of neurons were born within approximately ten minutes of each other.

Figure S2 shows the development of motor, sensory and interneurons. More than 80% of sensory and interneurons appear before hatching whereas this figure was found to be lower at around 50% for motor neurons. However, several of these motor neurons were polymodal, functioning also as inter-neurons. Only around 30% of exclusive motor neurons appeared before hatching. During such early development, there is less need for motor control: for example, earliest movements inside the shell do not involve neural coordination and first signs of neural activity are only observed around 30 minutes before hatching [12] .

**Analysis of spatial features – Connection-pairs**

The distance between any two connected neurons was calculated in three-dimensional space to determine the approximate length of the connection between them. Numerically, this was calculated as the three-dimensional Euclidean distance between two connected neurons and provided a useful measure to assess the length distribution of all edges. Although there is little information on the timing of synaptogenesis, we wanted to visualize how many of the neuron pairs forming short-, medium-, or long-distance connections in the adult were present at each stage. Hence, when we use the term '*connection pair*' - it does not imply synaptogenesis and is merely indicative of the birth of both neurons that will eventually connect. For the adult nematode, out of the 2,990 actual connections pairs 391 are long-distance (~ 14%), 298 are medium-distance (~ 10%) and 2,301 (~ 76%) are short-distance (see Methods for details).

Figure 4A shows the percentage of short, medium and long-range connection-pairs appearing at each stage of development. By the time of hatching (840 minutes), approximately 73% of short, 36% of medium and 68% of long-range connections-pairs had appeared, accounting for 69% of total connections pairs. For twenty random networks (Figure 4B) on the other hand, 52% (±3.5), 53% (±6.7) and 51% (±5.7) of short-, medium-, and long-distance connection-pairs respectively, occurred before hatching. In none of the cases did the actual value fall within the data domain of random tests. Further analysis revealed that the observed difference between the percentage of connection-pairs appearing before hatching in real and random systems was statistically significant, with $p < 0.001$ in each case.

To determine whether there was any relation between the degree of neuron and the proportion of short or long distance connections that they possessed, we computed the Pearson's coefficient for degree versus short and long connections. No significant correlation was found, with the correlation coefficient of degree with short and long connections being -0.03 and 0.11 respectively.

We then segregated the three types of junctions, namely, gap junction, chemical synapse or a combination of both. (If both gap junctions and chemical synapses existed between any two neurons, then the connection was termed as a combination.). The proportion of short-, medium- and long-length connections in each category is listed in Supplementary Table S1. Figure 5 compares the time of appearance of connection-pairs linked by gap junctions and chemical synapses in the adult, for short and long-distance connection lengths.  Around 35% of short- and 27% of long-distance connections in *C. elegans* are electrically coupled (Figure 5A), together constituting more than one-third of all connections in the nematode. Interestingly, electrically and chemically connected neuron pairs appear at the same rate with approximately 70% of each category appearing before hatching.

A bar chart was also plotted to visualize the distribution of long, medium, and short-distance connection-pairs during each of the developmental stages (Figure S3). Short distance connection-pairs are most abundant during all stages of development. The frequency of short-distance connections is followed by long-distance and then by medium-distance connection-pairs.

**Analysis of functional circuits**

The neurons of *C. elegans* have membership in various functional circuits [6] and we analyzed the appearance of connection-pairs in relation to this functional segregation. The percentage of each circuit that had appeared was measured in relation to the birth of connection-pairs. If a connection-pair involved neurons from two different circuits, then it was considered to belong to both circuits. Figure 6 shows how the various circuits appear over time. In the analysis of temporal features it was shown that approximately 80% of sensory and 50% of motor neurons appeared before hatching. Here, more specifically, analysis showed that the connection-pairs associated with circuitry of amphids, motoneurons in the nerve ring and other sensory receptors in the head, which are understandably more relevant to the early life of the worm are born sooner than other circuits like that of egg-laying, not required in the embryonic stages. Hence although we observe connection pairs belonging to all circuits present to various degrees in the earliest embryonic stages, the order of functional precedence may also influence the likelihood of early contact or interaction.

**Analysis of topological features of possible networks**

The networks at each stage were predicted based on the earliest possible time of synaptogenesis - when both neurons had appeared. Here, the network represents the neurons present at each stage with adult connectivity (see Methods). Our motivation behind this was to present the potential of network analysis in inferring development features like significant periods, from even discrete data. The global network over time was analyzed for topological changes such as number of nodes and edges (Figure S4A,B), clustering coefficient (Figure S4C,E), and characteristic path length (Figure S4D,F). The ratio of actual clustering coefficient to random of more than 2 and actual average path length to random of less than 1.5 are considered to signify a small-world network. The adult *C. elegans* network has already been shown to have small-world characteristics [13], here we find that all the networks display small-world characteristics with the ratio of actual clustering coefficient to that of random networks being above 4 at all times (Figure S4C), and the ratio of characteristic path length in the actual to the random network being consistently below 1.15 (Figure S4D). Although this analysis does not consider the effect of pruning, as the ratios are well beyond the characteristic ratios, the results are likely to be robust for small changes. The random networks created for these calculations were Erdős-Rényi networks which only maintained the total number of neurons and connections without preserving the degree distribution.

**Discussion**

Detailed analysis on the developing neural network in *C. elegans* revealed that timings of neurons as well as the connections between them were reflected in characteristic patterns of development. Based on the three-dimensional positions of the neurons and their known connectivity data, we showed that long-distance connection-pairs often appear in the early stages of development. At these times before hatching, neurons are nearby so that there is less need for long-distance cues and axon guidance. Finally, early-born neurons were highly connected (hubs) in contrast to neurons born later, which could be attributed to the longer time frame available to them to establish connections.

A significant proportion of connected neuron births were temporally close, with over half of the connected neurons appearing within an hour of each other. At the time of hatching (840 minutes), when the worm would be less than 20% of its final size [5], approximately 68% of long-length and 73% of short-length connection-pairs had already appeared. In contrast, random growth of a network of similar degree distribution as the actual network resulted in approximately half of all connection lengths appearing by the time of hatching. It may be noted that the connection length throughout this study refers to the spatial separation of the soma of the neurons that are connected and is not with respect to the location of synapse. The axonal length may be different from the overall connection length depending on the proximity or separation of the pre-synaptic neuron from the synapse or junction.

Temporal closeness in early development would imply spatial proximity, therefore reducing hurdles in wiring neurons that are far apart in the adult. It is therefore likely that early contact with subsequent axon extension will be preferred. This presumably would aid in minimizing metabolic costs sustained in wiring establishment during development, as axon outgrowth is an inherently more resource intensive pathway than axon extension [14,15]. Note that this cost minimization is only with respect to establishing the wiring. It should not be compared with sustained metabolic cost optimization through alternative wiring or cell migration costs that would be incurred irrespective of the time of synaptogenesis. The *C. elegans* neuronal network that has been the subject of several studies, is understood to be wired for efficiency of processing [2,10]. It has also been shown that while this spatial organization is not the one that minimizes total wiring length, the neurons are placed close to their optimal positions within the worm body for reducing wiring [16]. Here our results suggest that *C. elegans* has also evolved a pattern of development that enables it to establish this known connection pattern (particularly the substantial number of long-range connections) efficiently during growth.

The need for early contact has also been observed in other systems including the development of the excretory canal in *C. elegans* [17]. The canal tip is guided to and makes contact with the basement membrane early on in development. Subsequent growth involves a passive extension of the canal along the hypodermal membrane. It has been found that absence of this early contact, results in a slowing of growth in the larval stage, with the canal being unable to reach its target region [18]. The influence of early guidance has also been observed in the migration of the sex myoblast [18,19]. There are two mechanisms involved in guiding the sex myoblasts to the gonads, one acting early and the other much later, with absence of early guidance resulting in delays in the migration.

It is generally accepted that the most significant aspect of forming a synapse is contact between the neurons or neurites. While the role of pioneer axons has been studied in *C. elegans* [20], the fact that nearly 70% of connected neurons appear much before hatching, makes it more likely that cell-cell contact or interaction is established early on, particularly within circuits having functional significance in early life. There is evidence showing that several guidance molecules like Netrin and Nerfin-1 are expressed in the early stages of development. Netrin is a protein involved in axonal guidance in vertebrates as well as in invertebrates [21,22,23] and is specifically known to influence early path-finding events [23,24,25]. Nerfin-1 belonging to a highly conserved family of Zn-finger proteins, is found transiently expressed in neuron precursors and plays a role in early path finding. Studies involving pioneering neurons in the central nervous system of *Drosophila melanogaster* have shown that Nerfin-1, whose expression is spatially and temporally regulated [25], is essential in early axonal guidance. Disruption of genes involved in early guidance and migration could result in improper connectivity [26,27], as could delays in cell division. Our results suggest that neuron-neuron contact in early development could be important for achieving accurate connectivity. This would be significant not just in invertebrates but also in vertebrates where similar temporal expression profiles are seen for genes involved in early guidance. Inaccurately timed appearances of long-range connections or pioneer neurons would also influence subsequent connectivity. In addition, in *C. elegans*, as over two-thirds of long distance connection-pairs are born early on, it is likely that the interactions among the growing axons will play a role in fasciculation as observed in *Drosophila* [26].

The analysis of the formation of connection-pairs with respect to the nature of the junction has highlighted another interesting feature while also raising an important question. Approximately a third of all junctions are electrically coupled with around 70% each of long- and short-range connection-pairs appearing early in development. While little is known about the formation of gap junctions, they are most often seen in adjacent cells and hence it is likely that the 70% of connection-pairs appearing early will connect early. However, the interesting question raised here is of the remaining 30% of long-range connections coupled by gap junctions that appear in the later stages of development. Specifically, how are the neurites guided to their correct targets for the formation of gap junctions? The role of connexin proteins [28] and cell adhesion molecules like cadherins [29] in gap junction formation has been studied, however they would be insufficient to explain the formation of long-range axonal gap-junctions between neurons that are far apart. Theses gap junctions are mostly between late-forming ventral cord neurons and the existing pharyngeal ring neurons in the head. As the ventral cord neurons appear at later stages in development when the worm has already grown to approximately half of its final size, neuronal contact with subsequent migration is highly unlikely. While it is interesting to consider a dependence on chemically guided axons through fasciculation, in the absence of experimental data, these remain speculations.

Another important finding was that availability of time would be an important factor in the generation of network hubs. In the *C. elegans* network, nearly all neurons (~98%) with degree more than 30 were born before hatching, whereas only 74% (±5.5) would have occurred for random growth. Developmental time windows have been identified in the past [30] as playing a role in generating network hubs. A high probability in random networks, with an even higher value in the actual network, is indicative of the importance of time availability in establishing hubs. Indeed functional significance cannot be undermined and it needs to be emphasized that while being essential for achieving high number of connections, time is unlikely to be the deterministic factor attributing high degree connectivity to a particular neuron. Additionally, bilaterally symmetric neurons were born close to each other in time. Hence although time of birth is known to play an important role [31,32,33], in *C. elegans*, the asymmetry in function is less likely to be a result of environment-mediated change.

**Conclusion**

The neuronal network of *C. elegans*, being the only fully characterized connectome to date, has provided the opportunity to observe changes occurring during the course of its neural development. Based on available data, we have extracted time of creation of neurons to capture changes during growth and have identified features of neural development that would be significant in establishing long-range connections and network hubs. Continuous monitoring of synaptic connectivity during development is a steep challenge that goes beyond the current approaches for determining the adult connectome of different species [34,35]. We present an alternative, more feasible approach of employing global analysis on networks existing at discrete times of growth. With the aid of recent advances, obtaining connectivity information for these time stages could be possible in the near future. Availability of such data on connectivity will permit more detailed analyses, to give an insight into the structural changes unfolding during development. An interesting question that has been raised here is what mechanism ensures accuracy of wiring in late-forming, electrically coupled long-distance connections, and a clear answer is as yet unavailable. We hope that this work will stimulate further experimental and theoretical work on the network development of neural systems and *C. elegans* in particular.

**Materials and Methods**

We have produced a spatial representation of the neuronal network of *C. elegans* in three-dimensional space, so that the network resembled the anatomical network as much as possible. Three-dimensional coordinates were based on the two-dimensional spatial information of *C. elegans* neurons [36], which were updated with spatial information of three neurons that had been excluded. The data for neurons that did not have associated spatial information were obtained based on the spatial data of corresponding bilateral counterparts. The connectivity details from earlier studies [12,16,37] published in the Worm Atlas was used for the analysis. The ventral cord neuron VC6 that only makes connections through neuro-muscular junctions was not included here. Three loop connections (connections of a neuron to itself) were also excluded; as such connections did not influence our spatial and topological measures. Thus a total of 279 neurons and corresponding 2,990 connections were used. This included 1,584 uni-directional and 1,406 bi-directional connections. Biologically, they represent 672 gap junctions, 1962 chemical sysnapses and 376 connections where both gap junctions and chemical synapses exist between the neuron pairs. The latter were represented as bi-directional edges in the connection matrix. Neuro-muscular junctions were not included in the analysis. The coordinate information represents the positions of the soma of the neurons in three-dimensional space. The third dimension of each of the neurons was obtained by treating the body of the worm as a cylinder, guided by the actual shape of the worm. The third spatial coordinate for any neuron was then calculated as a function of the radius of the body of the worm and its known position along the y-axis, as follows:

$$z = \sqrt{r^2 - y^2}$$

where, $r$ – radius of the worm, was assumed to be constant along the length of the nematode (50 μm). This returned a positive value of $z$, and as many neurons in *C. elegans* have a left or right orientation, the physiological information available [5] was also used in determining the third coordinate. The three-dimensional coordinate was computed so that the spatial properties were closer to reality. Although as *C. elegans* has a very high length to diameter ratio, the results are unlikely to be affected even if the data had been two-dimensional. The left and right neurons were differentiated into positive and negative values, while those lying along the dorsal and ventral line had their third coordinate as zero.

To trace the growth of the network over time, we used the time estimates provided by Sulston *et al.* (1977 & 1983), in the cell lineage charts. The image files representing the lineage charts were read in and then analyzed to obtain the time of creation of each of the neurons. Based on this information, the neuronal network of *C. elegans* was obtained at different stages of growth. The margin of error in the embryonic lineage, as published, is 10% and 2% in the post-embryonic lineages. We produced spatial representations of the network at times of 350, 400, 500, 800, 2000 and 2700 minutes after fertilization. The embryo hatches at around 840 minutes [3], and the network at that time stage was found to be identical to that at 600 minutes in terms of neurons present. The choice of the time interval between the successive stages was based on the number of neurons appearing at different stages. It can be seen from the postembryonic lineage chart that there are very few neurons being created within first 20 hours after hatching (1200 minutes after hatching or 2000 minutes from fertilization). We therefore chose 2000 minutes post-fertilization as the next stage of development. As all but two neurons in *C. elegans* derive from the AB cell lineage, cross lineage comparisons were not performed. Although the network contained chemical synapses connecting one neuron to another as well as gap junctions coupling both neurons, the networks were treated as unweighted and directed as more than half of the connections (53%) were unidirectional. Gap junctions were represented as bi-directional. The node degree included both the incoming and outgoing connections.

Random networks created for comparative analysis had the same degree distribution as the actual *C. elegans* network. Neuron identities were randomly shuffled, so that all quantities estimated such as connection-length, birth-time difference, etc, would be modified. A neuron's identity referred to its spatial position and birth-time.

At any given time, a connection-pair existed if both neurons forming that connection (as in the adult) were present, without however, implying the formation of a synapse between them. The pair-wise, time-difference in the birth of neurons forming each of the 2990 connection-pairs was determined. The values were then separated into ten bins and mean of the time-differences in each bin was computed for enabling comparisons with random networks. The connection length between two connected neurons was the Euclidean distance between them in three-dimensional space. The lengths at each stage were separated into ten bins of size 0.12mm each, the first three were considered to be short distance (i.e. shorter than 0.36mm), the middle four as medium-range and the final three as long-range connections (i.e. longer than 0.84mm). This classification was used as the first three bins and the final three bins displayed a very high frequency of connections, whereas the intermediate bins were sparsely populated.

**Acknowledgements**

We thank Dr. David Hall for helpful comments on the manuscript.

**FIGURES**

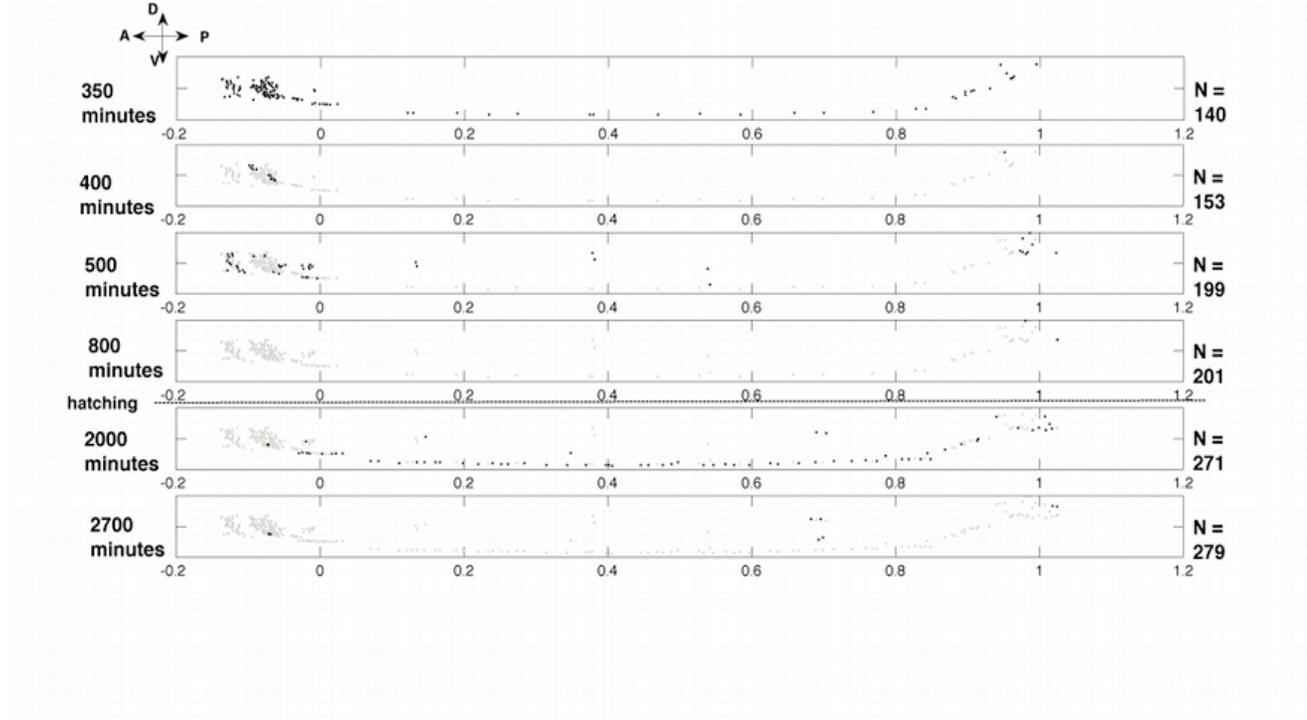

**Figure 1.** Growth of the neural network represented in six discrete time stages, providing a convenient visualization of the various neurons appearing before (gray) and during (black) the respective time stages. Orientations are along the Anterior (A)-Posterior (P), and Dorsal (D)-Ventral (V) axes. The horizontal dashed line indicates the approximate time of hatching (840 minutes). For any stage, the dots in grey are the neurons that have already appeared, while the black dots indicate those that appear by the end of that particular stage.

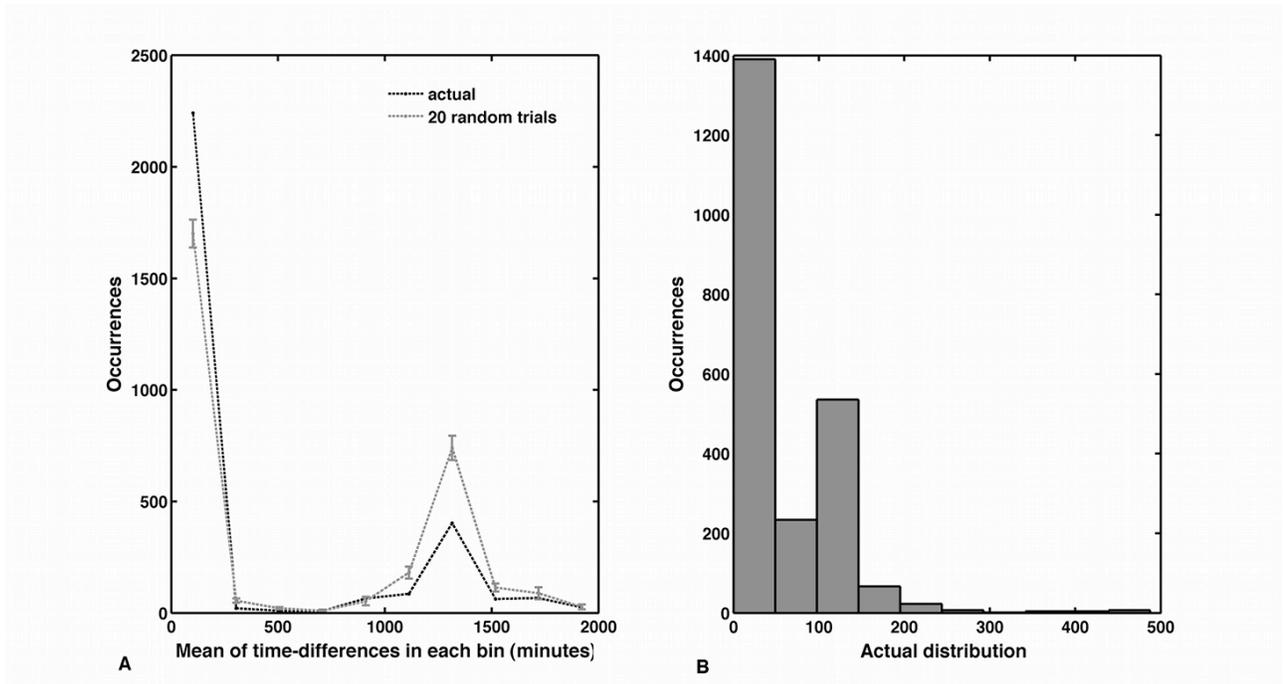

**Figure 2.** Time difference in births of connected neurons. Most connected neurons are born temporally close to each other with approximately half of them appearing within 50 minutes of each other. Comparison shows that similar findings cannot be expected from random growth. (A) Mean of birth-time-differences of all pairs of connected neurons plotted against the frequency within each time interval. Mean of time-differences refers to the mean of the time differences within each bin. Curves show the actual network compared to average values and standard deviations from twenty random networks. (B) The actual distribution of the birth-time-differences within the first ten bins giving a more detailed picture of the most prevalent birth-time-differences between connected neurons. The first 50 minutes bin has the highest frequency at approximately 1400. Thus around 1400 connected neurons (~half of all connection-pairs) form within 50 minutes of each other.

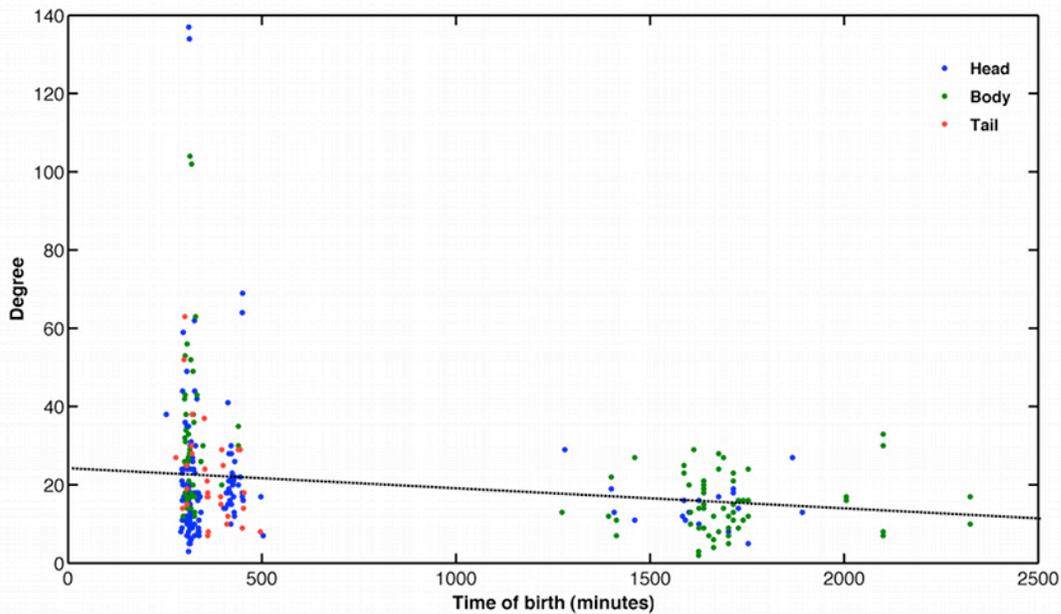

**Figure 3.** Node degree, the number of connections of a neuron, in relation to its time of origin and position in the nematode body, namely, Head, Body or Tail. Note that all nodes with more than 33 edges originate early during development. As would be expected the highest connectivity neurons are in the head. The dashed line represents the regression for a linear fit to the data. (Pearson's correlation coefficient R = -0.24).

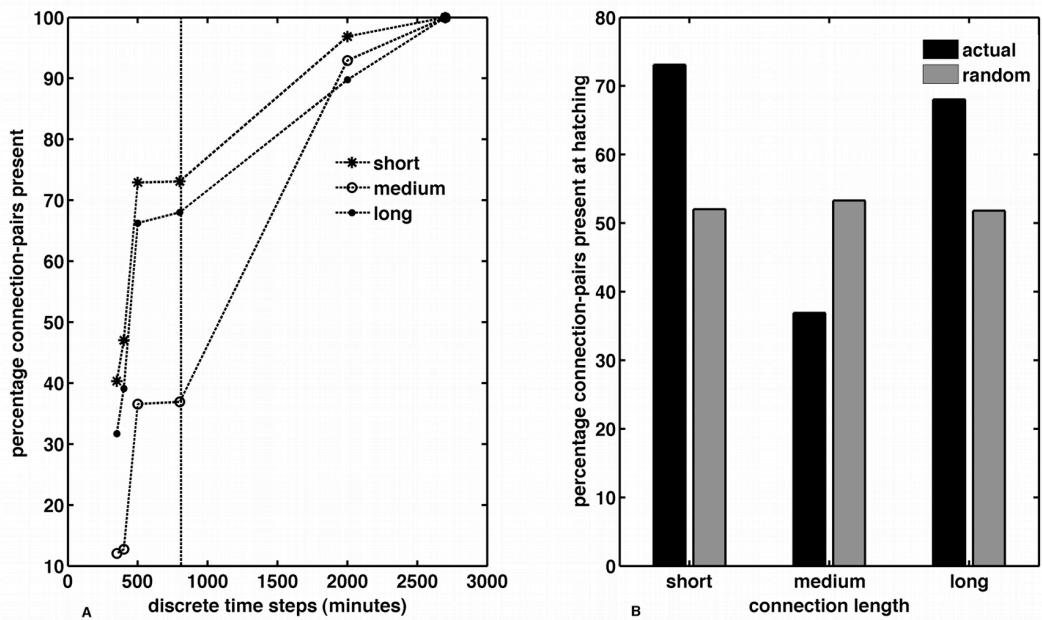

**Figure 4.** Variation in the growth of short, medium and long-range connection-pairs during development, compared against random networks (discrete time steps). (A) Percentage of each type of connection-pairs appearing during development. Percentage is with respect to the total number of connections of that type (short, medium or long) in the adult. The black vertical dashed line represents the time of hatching. (B) Comparison with random networks, of the percentage of each category of connection-pairs appearing before hatching.

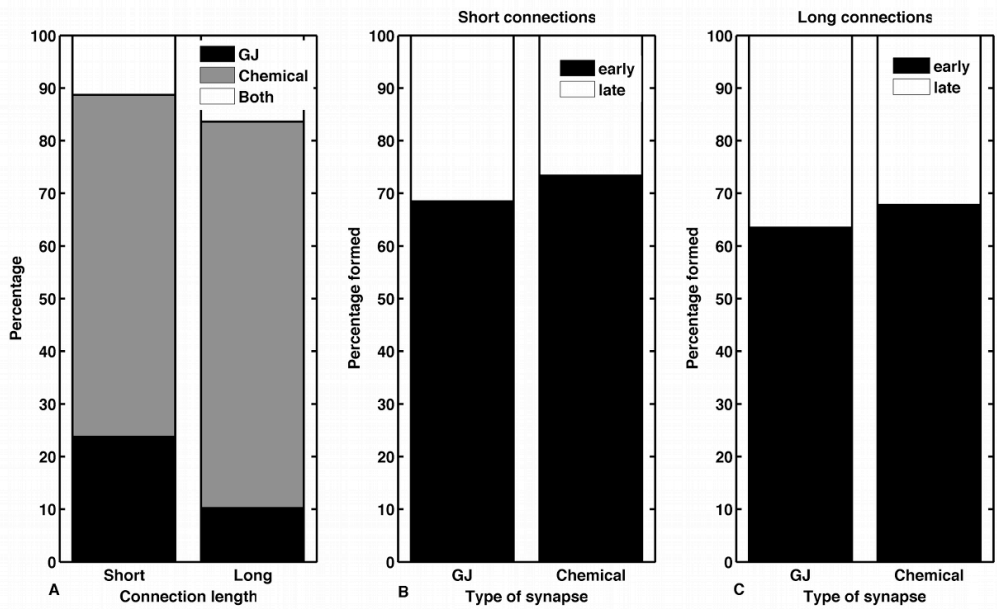

**Figure 5.** Analysis of gap junctions and chemical synapses in short and long range connections. Approximately a quarter of long-distance and a third of short-distance connections are coupled by gap junctions (including combination junctions). Of these around 70% appear before hatching. (A) Prevalence of gap junctions, chemical synapses and combinations of the two, in short and long distance connections. Appearance of connection-pairs linked by gap junctions and chemical synapses, for short (B) and long-length connections (C). Early and late respectively refer to time periods before and after hatching. (GJ: Gap Junction and Chemical: Chemical synapse)

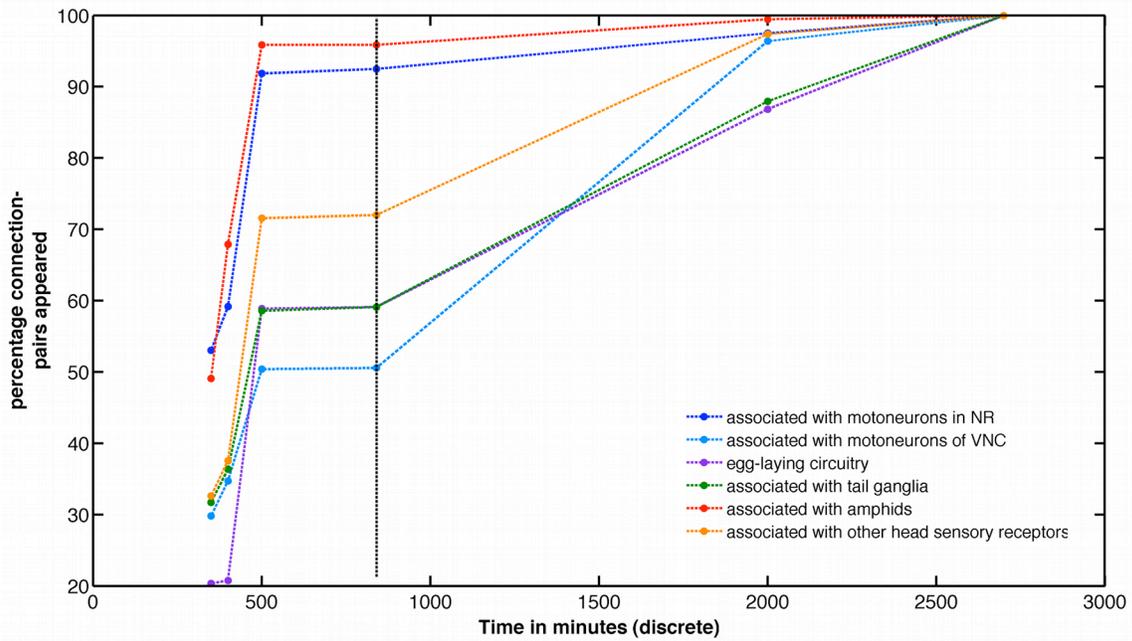

**Figure 6.** The progressive appearance of 'connection-pairs' belonging to the various circuits during development. Time along x-axis is represented in discrete time steps. The circuits associated with sensory feedback and survival appear sooner. Note that appearance is with respect to both neurons that are connected in the adult being born at that time, actual synaptogenesis may occur at a later time. The dashed line represents the time of hatching. (NR – Nerve Ring and VNC – Ventral Nerve Cord)

**Supplementary Figures and Legends**

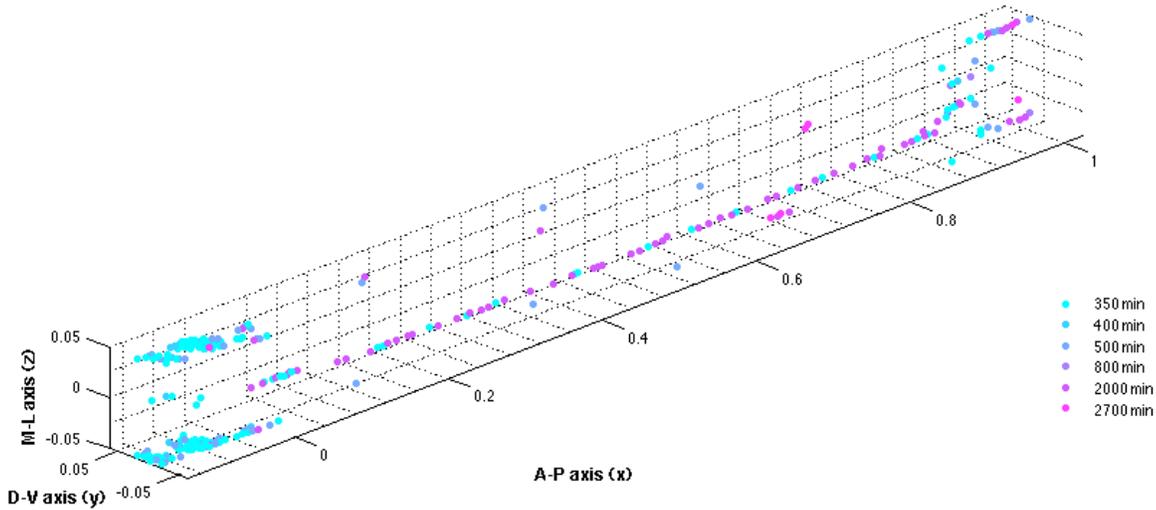

**Figure S1.** Neurons formed at successive stages of growth (color scale). Neuron positions are shown at the adult stage (x: anterior-posterior axis, y: dorso-ventral axis, z: medial-lateral axis). Note that some ventral cord neurons are established early whereas most occur at late stages of development.

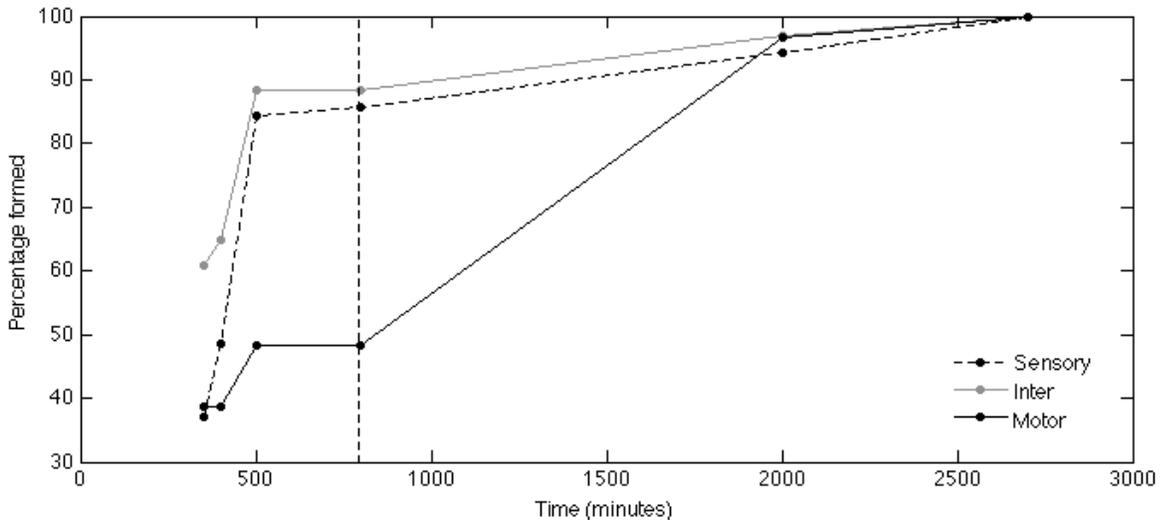

**Figure S3.** Growth pattern of sensory, motor and inter neurons. The vertical dashed line indicates the approximate time of hatching. The plot shows the percentage of neurons formed at each time step for the three neurons types (sensory, inter, and motor neurons). Note that while sensory and inter-neurons are formed early on, less than fifty percent of the motor neurons are formed at the time of hatching (840 minutes). By excluding polymodal neurons, the percentage drops further to around 30%. Motor neurons or sensory neurons that also function as interneurons constitute polymodal neurons.

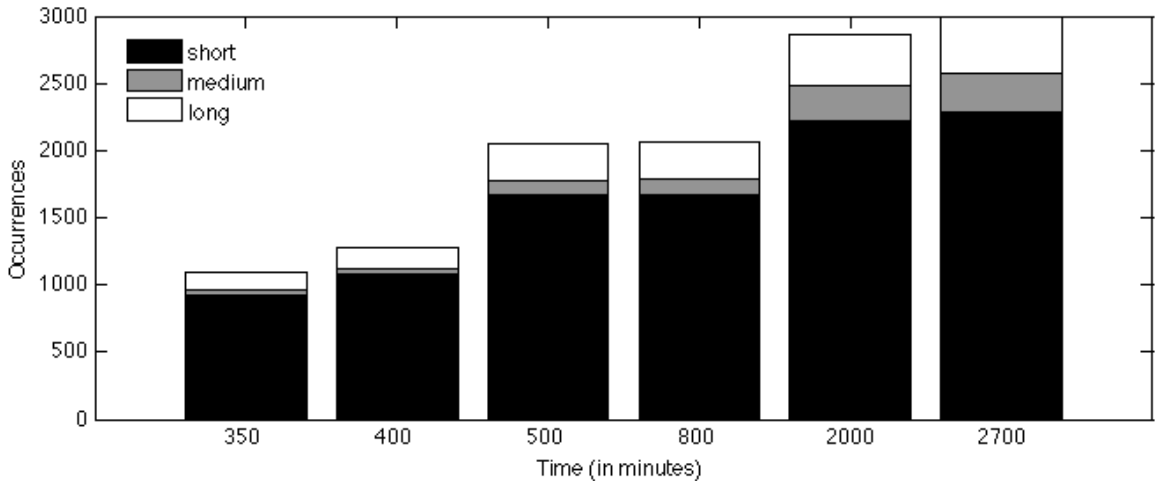

**Figure S4.** Distribution of approximated metric connection lengths during development as a reflection of neuron births. The metric lengths correspond to the adult and the presence at a particular development stage signifies the existence of both neurons that are connected in the adult. The proportion of small-, medium-, and long-distance connection pairs are shown in each bar and identified in the legend (See also Figure S1).

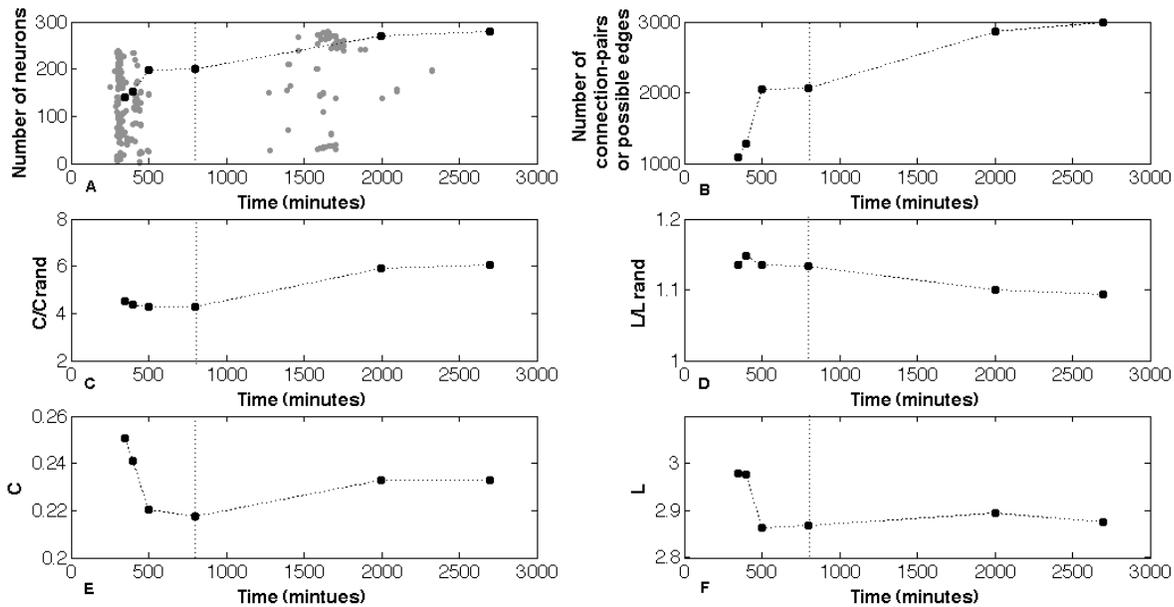

**Figure S5.** Change in topological network organization during development represented at each discrete time stage. (A) The total number of neurons formed by the end of each of the discrete time steps (black dots) superimposed on the actual birth times of neurons (gray dots) (C) Ratio of the clustering coefficient of the actual network compared to a random network ($C/C_{rand}$). (D) Ratio of the characteristic path length of the actual network compared to a random network ($L/L_{rand}$). (E) and (F) absolute values of the clustering coefficient $C$ and characteristic path length $L$, respectively (See also Figures S5 and S6). Time of hatching (approximately 840 minutes) is soon after the third discrete time step (800 minutes).

|  | Short | Medium | Long | Total |
| --- | --- | --- | --- | --- |
| Gap-junction | 546 | 66 | 40 | 652 |
| Chemical synapse | 1495 | 180 | 287 | 1962 |
| Combination | 260 | 52 | 64 | 376 |
| Total | 2301 | 298 | 391 | 2990 |

Table S1. Short, medium and long-length connections separated by type of synapse.